\documentclass[prl,aps,10pt,twocolumn,showkeys,showpacs,nofootinbibs]{revtex4-1}

\usepackage{amsmath,amssymb,amsfonts,amsthm}
\usepackage{amsbsy} 
\usepackage{epsfig}
\usepackage{latexsym}
\usepackage{color}
\usepackage[utf8]{inputenc}

\usepackage{graphicx}
\usepackage{caption}
\usepackage{subcaption}

\usepackage{hyperref}

\def\barr{\begin{array}}
\def\earr{\end{array}}

\def\ben{\begin{equation}}
\def\een{\end{equation}}
\def\bs{\begin{subequations}}
\def\es{\end{subequations}}
\def\bena{\begin{eqnarray}}
\def\eena{\end{eqnarray}}

\def\be{\begin{equation}}
\def\ee{\end{equation}}

\def\bes{\begin{eqnarray}}
\def\ees{\end{eqnarray}}

\begin{document}

\title{Shocks in the Early Universe}

\author{Ue-Li Pen }
\affiliation{Canadian Institute for Theoretical Astrophysics, 60 St George St, Toronto, ON M5S 3H8, Canada}

\author{Neil Turok}
\affiliation{Perimeter Institute for Theoretical Physics, Waterloo ON N2L 2Y5, Canada}

\date{\today}


\begin{abstract}
We point out a surprising consequence of the usually assumed initial conditions for cosmological perturbations. 
Namely, a spectrum of Gaussian, linear, adiabatic, scalar, growing mode perturbations not only creates acoustic oscillations of the kind observed on very large scales today, it also leads to the production of shocks in the radiation fluid of the very early universe. Shocks cause departures from local thermal equilibrium as well as creating vorticity and gravitational waves. For a scale-invariant spectrum and standard model physics, shocks form for temperatures $1$ GeV$<T<10^{7}$ GeV. For more general power spectra, such as have been invoked to form primordial black holes, shock formation and the consequent gravitational wave emission provides a signal detectable by current and planned gravitational wave experiments, allowing them to strongly constrain conditions present in the primordial universe as early as $10^{-30}$ seconds after the big bang. 
\end{abstract}
\maketitle

Over the past two decades, observations have lent powerful support to a very simple model of the early universe: a flat, radiation-dominated Friedmann-Lema\^{i}tre-Robinson-Walker (FLRW) background cosmology, with a spectrum of small-amplitude, growing perturbations. In this Letter we study the evolution of these perturbations on very small scales and at very early times.  The simplest and most natural possibility is that their spectrum was almost scale-invariant, with the rms fractional density perturbation $\epsilon \sim10^{-4}$ on all scales. However, more complicated spectra are also interesting to consider. For example, LIGO's recent detection of  $\sim 30 M_{\odot}$  black holes~\cite{LIGO} motivated some to propose a bump in the primordial spectrum with $\epsilon \sim 10^{-1}$ on the relevant comoving scale. High peaks on this scale would have collapsed shortly after crossing the Hubble horizon, at $t\sim 10^{-4}$ seconds, to form $30 M_{\odot}$  black holes in sufficient abundance to constitute the cosmological dark matter today~\cite{Bird}. 

Here we focus on the evolution of acoustic waves inside the Hubble horizon. In linear theory, they merely redshift away as the universe expands. However, higher order calculations~\cite{Gielen} revealed that perturbation theory fails due to secularly growing terms. We explain this here by showing, both analytically and numerically, in one, two and three dimensions, that small-amplitude waves steepen and form shocks, after $\sim \epsilon^{-1}$ oscillation periods~\cite{earlierrefs}: for movies and supplementary materials see \cite{weblink}. 

Furthermore, shock collisions would generate gravitational waves. As we shall later explain, the scenario of Ref.~\cite{Bird}, for example, would produce a stochastic gravitational wave background large enough to be detected by existing pulsar timing array measurements. More generally,  planned and future gravitational wave detectors will be sensitive to gravitational waves generated by shocks as early as  $10^{-30}$ seconds after the big bang~\cite{penturoklong}. 

\begin{figure}[htp]
\vskip -.6in
\includegraphics[scale=0.32]{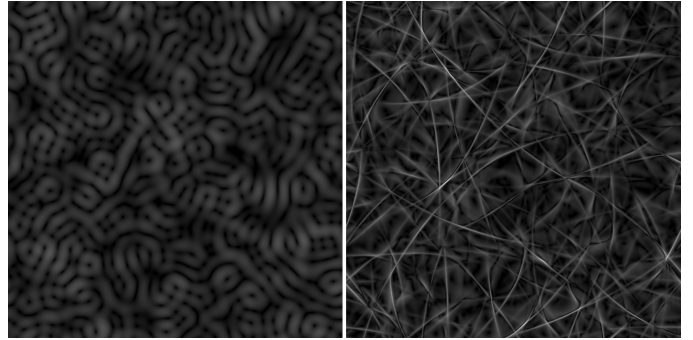}
\vskip -.5in
\caption{ \small Simulation showing cosmological initial conditions (left) evolving into shocks (right).  The magnitude of the gradient of the energy density is shown in greyscale. The initial spectrum is scale-invariant and cut off at ${1\over 128}$ the box size, with rms amplitude $\epsilon=.02$. Movie available at  \cite{weblink}.}
\label{fig:movie}
\vskip -.1in
\end{figure}

Shock formation also has important thermodynamic consequences. In a perfect fluid, entropy is conserved and the dynamics is reversible. The presence of a spectrum of acoustic modes means that the entropy is lower than that of the homogeneous state but, within the perfect fluid description, the entropy cannot increase. Shock formation leads to the breakdown of the fluid equations, although the evolution is still determined by local conservation laws. Within this description, shocks generate entropy, allowing the maximum entropy,  thermal equilibrium state to be achieved. Shock collisions also generate vorticity, a process likewise forbidden by the fluid equations. Both effects involve strong departures from local equilibrium and are of potential relevance to early-universe puzzles including the generation of primordial magnetic fields and baryogenesis~\cite{penturoklong}. 

Of course, the perfect fluid description is not exact and dissipative processes operate on small scales.  In fact, the shock width $L_s$ is set by the shear viscosity $\eta$, and the density jump $\epsilon\, \rho$ across the shock. For a relativistic equation of state, {\it i.e.}, $P=c_s^2 \rho$, with $c_s=1/\sqrt{3}$, we find $L_s=9\sqrt{2}\,\eta/(\epsilon \,\rho)$ ~\cite{LL,penturoklong}. For shocks to form, $L_s$ must be smaller than the scale undergoing non-linear steepening, of order $\epsilon$ times the Hubble radius $H^{-1}$. In the standard model, at temperatures above $\sim 100 $ GeV, the right handed leptons, coupling mainly through weak hypercharge, dominate the viscosity, yielding $\eta\sim 16 /g'^4\ln(1/g') \sim 400 \,T^3$~ \cite{moore}. Using $\rho=(\pi^2/30){\cal N} T^4$, with ${\cal N}=106.75$,  we find $L_s$ falls below $\epsilon H^{-1}$ and hence shocks form when $T$ falls below $\sim \epsilon^2 10^{15}$ GeV, {\it i.e.}, $10^7$ GeV for $\epsilon=10^{-4}$.  At the electroweak temperature viscous effects are negligible both in shock formation and, as we discuss later, shock decay.  However, once $T$ falls below the electroweak scale, the Higgs field gains a vev $v$ and the neutrino mean free path grows as $\sim v^4/T^5$, exceeding $10^{-4} H^{-1}$ when $T$ falls below $\sim 1$ GeV for $\epsilon=10^{-4}$ or $\sim 100$ MeV for  $\epsilon=10^{-1}$ . At lower temperatures,  acoustic waves are damped away by neutrino scattering before they steepen into shocks.

\begin{figure}[htp]
\includegraphics[scale=0.35]{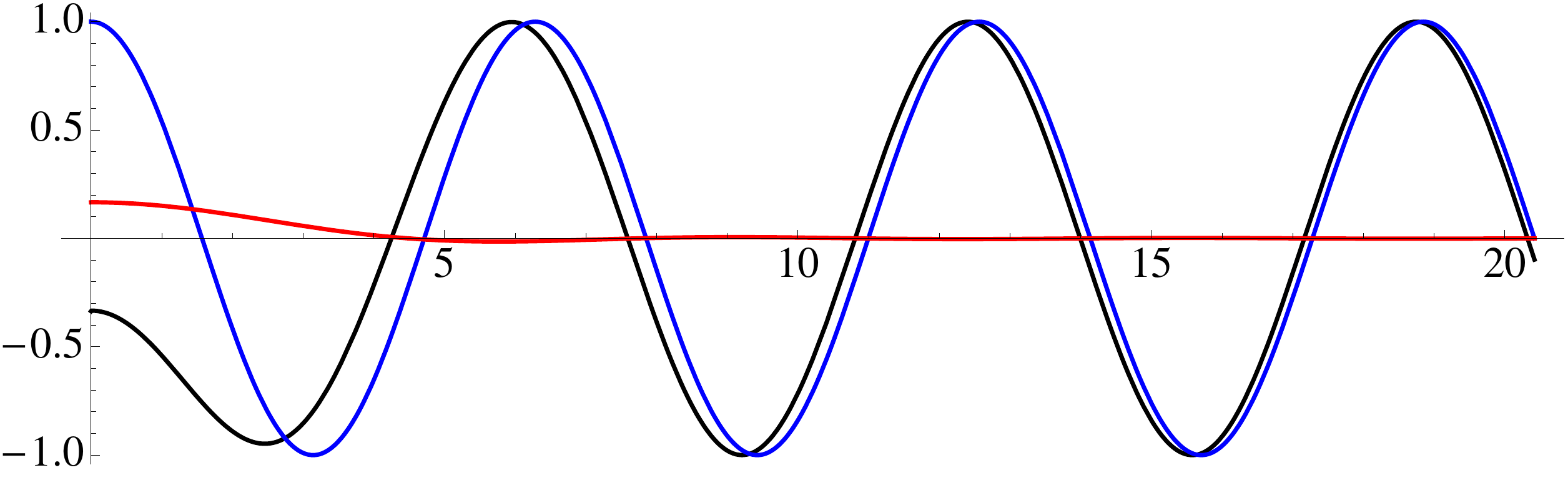}
\caption{The growing mode perturbation, in a radiation-dominated universe, in conformal Newtonian gauge. The density perturbation $\delta_k(t)$ (black), the  Newtonian potential $\Phi$ (red) and the flat spacetime approximation to $\delta_k(t)$ (blue) are plotted against $kc_s t $.}
\label{modesfig}
\end{figure}

This Letter is devoted to the early, radiation-dominated epoch in which shocks form. We assume standard, adiabatic, growing mode perturbations. Their evolution is shown in Fig. \ref{modesfig}: as a mode crosses the Hubble radius, the fluid starts to oscillate as a standing wave, and the associated metric perturbations decay. Thereafter, the fluid evolves as if it is in an unperturbed FLRW background. The tracelessness of the stress-energy tensor means that the evolution of the fluid is identical, up to a Weyl rescaling, to that in Minkowski spacetime, where the conformal time and comoving cosmological coordinates are mapped to the usual Minkowski coordinates. 

In flat spacetime,  the fluid equations read $\partial_\mu T^{\mu \nu}=0$. For a constant equation of state, $P=c_s^2 \rho$, we have $T^{\mu \nu}=(1+c_s^2) \rho \, u^\mu u^\nu +c_s^2 \rho\, \eta^{\mu \nu}$, with $u^\mu=\gamma_{v}(1,\vec{v})$ the fluid 4-velocity. In linear theory, the fractional density perturbation $\delta$ and velocity potential $\phi$ (with $\vec{v}= \vec{\nabla} \phi$) obey the continuity equation $\dot{\delta}=-(1+c_s^2) \vec{\nabla}^2 \phi$ and the acceleration equation $\dot{\phi}=-c_s^2/(1+c_s^2)\, \delta$. Setting $\delta(t,\vec{x})=\sum_{\vec{k}} \delta^{(1)}_{\vec{k}} (t) e^{i\vec{k}.\vec{x}}$,  for scale-invariant, Gaussian cosmological perturbations on sub-Hubble scales the statistical ensemble is completely characterized  by
\ben
\langle\delta^{(1)}_{\vec{k}}(t)\delta^{(1)}_{\vec{k}'}(t') \rangle=\delta_{\vec{k}+\vec{k}',\vec{0}}\,{2 \pi^2 {\cal A}\over k^3 V} \cos(k c_s t) \cos(k c_s t')
\label{eq1}
\een
where ${\cal A} \equiv \epsilon^2$ is the variance per log interval in $k$ and $V$ is a large comoving box. From Planck measurements, we determine $\epsilon\approx 6\times 10^{-5}$~\cite{foottilt}.

{\it Wave steepening:} 
The fluid energy-momentum tensor $T^{\mu \nu}$ depends on four independent variables, $\rho$ and $\vec{v}$. So the spatial stresses $T^{ij}$ may be expressed in terms of  the four $T^{\mu0}$ and the four equations, $\partial_\mu T^{\mu \nu}=0$ used to determine the evolution of the fluid. For small-amplitude perturbations, we expand in $T^{0i}/\overline{T^{00}}$, where bar denotes spatial average, obtaining $T^{ij}\approx c_s^2 T^{00}\delta^{ij} +(T^{i 0}T^{j 0}-c_s^2\delta^{ij} T^{0k} T^{0k})/((1+c_s^2)\overline{T^{00}})$ at second order.

At the linearized level, a standing wave is the sum of a left-moving and a right-moving wave. Assuming planar symmetry, we define $\Pi\equiv T^{01}/\overline{T^{00}}$. Consider a right-moving linearized wave $\Pi^{(1)}(u)$, where $u\equiv x-c_s t$, $v\equiv x+c_s t$. To second order, the fluid equations read $2\kappa \partial_u \partial_v \Pi +(\partial_u^2-\partial_v^2)\Pi^2=0$, with $\kappa=2 c_s(1+c_s^2)/(1-c_s^2)$. For the given initial condition, $v$ derivatives are suppressed relative to $u$ derivatives by one power of  $\Pi$. Hence we can drop the $\partial_v^2\Pi^2$ term and integrate once in $u$ to obtain Burger's equation,
\be
\kappa \, \partial_v \Pi + \Pi \partial_u \Pi=0,
\label{eq2}
\ee
for which, as is well-known, generic smooth initial data $\Pi(0,u)$ develop discontinuities in finite time $v$. 

Equation (\ref{eq2}) may be solved exactly by the method of characteristics: the solution propagates along straight lines, so that $\Pi(v,u+\Pi(0,u) v/\kappa)=\Pi(0,u)$, where $\Pi(0,u)$ is initial data at $v=0$. Consider a standing wave $\delta=-\sqrt{2} \epsilon \sin k x \cos k c_s t$, with initial variance $\epsilon^2$. Decomposing it into left- and right-moving waves, the latter is $\delta^{(1)}=- \epsilon \sin(k u)/\sqrt{2} $ and, correspondingly, $\Pi^{(1)}=- c_s \epsilon \sin (k u)/\sqrt{2}.$ The characteristic lines first intersect at $u=0$ and $v=\sqrt{2}\kappa/(k c_s\epsilon)$, {\it i.e.}, when $t=\kappa /(k c^2_s \epsilon)$. Setting $c_s=1/\sqrt{3}$, we conclude that shocks form when $k c_s t \epsilon \sim \sqrt{8}$ or
after $\sqrt{2}/ (\pi \epsilon)$ oscillation periods. The wave steepening effect is also seen in perturbation theory.  From (\ref{eq2}) one finds $\Pi^{(2)}=-c_s^2 \epsilon^2 (k v/4\kappa) \sin (2 k u)$,  steepening $\Pi$ around its zero at $u=0$, with the second order contribution to the gradient equalling the first order contribution precisely when $k c_s t \epsilon \sim \sqrt{8}$.

{\it Characteristic rays:} In higher dimensions, we can likewise gain insight into shock formation by following characteristic rays. These are the trajectories followed by small amplitude, short-wavelength disturbances~\cite{LL}, moving in the background provided by the perturbed fluid. For a perfect fluid with $c_s=1/\sqrt{3}$, if the 3-vorticity $\vec{\nabla}\wedge (\rho^{1\over 4} \vec{u})$ is initially zero, it remains zero for all time. We can then write $\rho^{1\over 4} \vec{u} = \overline{\rho}^{1\over 4} \vec{\nabla} \phi$, with $\phi$ a potential, at least until shocks form. We write the perturbed density and potential as: $\rho=\overline{\rho}(1+\delta_b+d \delta)$ and  $\phi=\delta \phi_{b}+ d\phi$, where $\delta_{b}$ and $\delta \phi_{b}$ represent a background of linearized waves and $d \delta$ and $d \phi$ represent short-wavelength disturbances.  The  evolution of $d \delta $ and $d \phi$ is governed by the second order perturbation equations, $\partial_t d \delta+{4\over 3} \vec{\nabla}^2 d \phi +{1\over 3} \vec{\nabla}\cdot (\delta_{b}\vec{\nabla}d\phi+d \delta \,\vec{\nabla}\phi_{b} )=0$ and $\partial_t d \phi +{1\over 4} d \delta  - {1\over 16} \delta _{b} d\delta+\vec{\nabla}\phi_{b}\cdot\vec{\nabla} d\phi=0$. These may be solved in the stationary phase approximation: we set $d\phi=A_\phi e^{i {S}}$ and $d\delta=A_\delta e^{i {S}}$ and assume that $A_\phi$ and $A_\delta$ vary slowly so that the variation of the phase $S$ controls the wave fronts. The leading (imaginary) part of the equations of motion yields a linear eigenvalue problem for $A_\phi$ and $A_\delta$, with $i\partial_t S$ the eigenvalue.  We obtain
\ben
\partial_t  { S}=-{\sqrt{\vec{(\nabla}S)^2}\over \sqrt{3}}-{2\over 3} (\vec{\nabla}S \cdot\vec{\nabla}\phi_{b}),
\label{eq4}
\een
the Hamilton-Jacobi equation for a dynamical system with Hamiltonian ${\cal H}(\vec{p},\vec{x},t) =-\partial_t  {S}(t,\vec{x})$, where ${ S}(t,\vec{x}(t))$ is the action calculated on a natural path, {\it i.e}, a solution of the equations of motion. The Hamiltonian
\ben
{\cal H}(\vec{p},\vec{x},t) ={|\vec{p}\,|\over \sqrt{3}}+{2\over 3} \vec{p} \cdot\vec{\nabla}\phi_{b}(t,\vec{x})
\label{eq5}
\een
and the ray trajectories $\vec{x}(t)$ obey Hamilton's equations:
\ben
\dot{\vec{x}}={\vec{n}\over \sqrt{3}} +{2\over 3} \vec{\nabla}\phi_{b}, \quad \dot{n_i}=-{2\over 3} (\partial_i-n_i(n_j\partial_j))(\vec{n}\cdot \vec{\nabla}\phi_{b}),
\label{eq6}
\een
where $\vec{n}\equiv \vec{p}/|\vec{p}|$. Note that (due to scale invariance) ${\cal H}$  is homogeneous of degree unity in $\vec{p}$. It follows that (i) the ray velocities $\dot{\vec{x}}$ depend only on the direction of $\vec{p}$, not its magnitude, and (ii) the phase of the wave on the stationary-phase wavefront,  ${\cal S}=\int dt( \vec{p}\cdot\dot{\vec{x}} -{\cal H}),$ is a constant as a consequence of Hamilton's equations. Hence, when characteristic rays cross there are no diffractive or interference phenomena.

{\it Caustics and Shocks:} 
The set of all characteristic rays is obtained by solving the equations of motion (\ref{eq6}), for all possible initial positions and directions, $\vec{q}\equiv \vec{x}(0)$ and $\vec{m}\equiv \vec{n}(0)$. The solutions provide a mapping from $(\vec{q},\vec{m})$ to $(\vec{x},\vec{n})$, at each time $t$, which can become many to one through the formation of caustics~\cite{Arnold}. If so, the solution to the fluid equations can be expected to acquire discontinuities such as shocks.  The signature of the mapping becoming many to one is the vanishing of the Jacobian determinant $J\equiv|\partial (\vec{x}, \vec{n})/\partial (\vec{q} ,\vec{m})|$ at some $(\vec{q},\vec{m})$. We compute the change in this determinant in linear theory, and extrapolate to determine when it might vanish. The dominant effect comes from the deviation in the ray position which grows linearly in time (whereas the deviation in the ray direction does not). Thus we may approximate $\delta J \approx \delta |\partial \vec{x}/\partial \vec{q} |$. Setting $\vec{x}(t,\vec{q})=\vec{x}_0(t)+\vec{\psi}(t, \vec{q})$, where $\vec{x}_0(t)\equiv \vec{q}+\vec{m}\,t/\sqrt{3}$ is the unperturbed trajectory and $\vec{\psi}(t,\vec{q})$ is the displacement, we integrate (\ref{eq6}) in the approximation that $\psi$ is small, so that the spatial argument of $ \phi_b$ may be taken as $\vec{x}_0(t)$. To first order in $\psi$, $\delta J \approx \vec{\nabla}_{\vec{q}} \cdot \vec{\psi}(t,\vec{q})$. A rough criterion for $J$ to develop zeros and thus for shocks to form, in abundance, is that the variance $\langle(\delta J)^2\rangle$, computed in the Gaussian ensemble of linearized perturbations for $\phi_b$, attains unity. 

In these approximations, from (\ref{eq6}) we obtain
\ben
\delta J\approx {2\over 3} \int_0^t dt' \left(\vec{\nabla}_{\vec{q}}^2 -{t-t' \over \sqrt{3}} \,\hat{ O} \right) \phi_{b}(t',\vec{x}_0(t')),
\label{eq7}
\een
where $\hat{ O}\equiv \left(\vec{\nabla}_{\vec{q}}^2-(\vec{m}\cdot\vec{\nabla}_{\vec{q}})^2\right)(\vec{m}\cdot \vec{\nabla}_{\vec{q}})$.  The term involving $\hat{O}$ (which only exists for $d>1$) dominates at large $t$. It describes how gradients in the background fluid velocity deflect the ray direction $\vec{n}$, with each ``impulse'' on $\vec{n}$ contributing a linearly growing displacement to $\vec{\psi}$. 

We compute the variance $\langle (\delta J)^2 \rangle$ from (\ref{eq7}) by taking the ensemble average using the $\phi_b$ correlator implied by (\ref{eq1}). The contribution of modes with $k<k_c$ is given by 
\bena
 \langle (\delta J)^2\rangle \approx\left(k_c c_s t \,\epsilon \right)^2\times
  \begin{cases}
   {3\over 32}  & \quad  {\rm for} \quad d=1    \cr
  {1\over 16}     &\quad {\rm for} \quad d=2    \cr
  {1\over 24}     & \quad  {\rm for} \quad d=3,   \cr
  \end{cases}
\label{eq8}
\eena
so that, for example, in the 3d ensemble, at any time $t$ shocks form on a length scale $\lambda_s\approx (\pi/\sqrt{6}) \epsilon \, c_s t$.

{\it Simulations} We have implemented a fully relativistic TVD hydro code to solve the non-linear conservation equations in 1, 2 and 3 dimensions (always using $c_s=1/\sqrt{3}$). The code is a slight modification of \cite{trac} to relativistic fluids, and parallelizes on a single node under OpenMP.  For the initial conditions,  $T_{00}$ was taken to be perturbed with a scale-invariant Gaussian random field, and $T^{0i}$ was set zero, consistent with cosmological initial conditions.  The initial power was truncated at $N$ times the fundamental mode where, for example, $N=128$ for $1024^3$ simulations in 3d and $N=256$ for $4096^2$ simulations in 2d. Various initial perturbation amplitudes were simulated in order to check consistency with the analytical discussion provided above and below. Further details and movies of the simulations are provided at \cite{weblink}.

{\it Thermalization:} Consider the effect of an initially static density perturbation, $\rho(\vec{x})\rightarrow \overline{\rho}\left(1+\delta_i(\vec{x})\right)$, where $\overline{\rho}$ is the mean energy density. The fluid energy density is $T^{00}= {4\over 3} \rho \gamma_{v}^2-{1\over 3} \rho$, where $\vec{v}$ is the fluid velocity.  Expanding to quadratic order in the perturbations, we find $T^{00}(\vec{x})=\overline{\rho}(1+\delta+{4\over 3} \vec{v}^2)$. At the initial moment, $\vec{v}(x)$ is zero everywhere and the spatial average  $\overline{\delta_i}$ is zero by definition, hence $\overline{T^{00}} =\overline{\rho}$. However, once $\delta$ starts oscillating, a virial theorem holds, connecting the average variances: $\langle \vec{v}^2\rangle ={3\over 16}\langle  \delta^2\rangle$. Thus, energy conservation implies that $\overline{\delta}$ falls by ${1\over 4} \langle \delta^2 \rangle$, to compensate for the kinetic energy in the oscillating modes. The system is not, however, in local thermal equilibrium. The entropy density is given, up to a constant, by $\rho^{3\over 4} \gamma_{v}\approx \overline{\rho}^{3\over 4} (1+{3\over 4} \delta -{3\over 32} \delta^2 +{1\over 2} \vec{v}^2),$ to second order in the perturbations. Using energy conservation and the virial theorem, the fractional deficit in the mean entropy density is thus $-{3\over 16} \langle \delta^2 \rangle=-{3\over 32} \langle \delta_i^2 \rangle$, where $\delta_i$ is the initial density perturbation. For a scale-invariant spectrum of initial perturbations, the fractional entropy deficit contributed by waves of wavelengths $\lambda_{1}<\lambda<\lambda_{2}$ is  $-{3\over 32} \epsilon^2 \int_{\lambda_{1}}^{\lambda_{2}}(d\lambda/\lambda)=-{3\over 32} \epsilon^2\ln(\lambda_{2}/\lambda_{1})$.

 \begin{figure}[htp]
\vskip -.5in
\hskip-.42in
\includegraphics[scale=0.44]{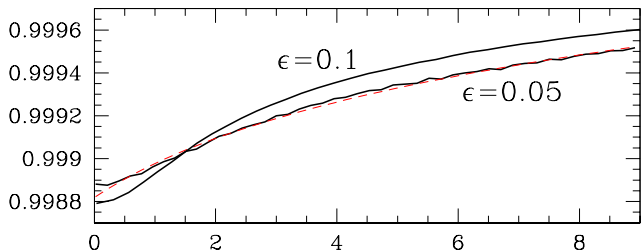}
\vskip -3in
\caption{Entropy, in units of its equilibrium value, versus the time $t$, in units of the sound-crossing time, for $512^3$ simulations of a perfect radiation fluid with cosmological initial conditions as in (\ref{eq1}). The red dashed curve is a fit to the $\epsilon=.05$ curve using (\ref{eq9}) with $C={1\over 4}$. For $\epsilon=0.1$, $t$ has been doubled and the entropy deficit rescaled by a quarter to verify good agreement with (\ref{eq9}).  }
\label{entropyfig}
\end{figure}

Once shocks form, they generate entropy at a rate which may be computed as follows~\cite{LL2}. Local energy-momentum conservation requires that the incoming and outgoing energy and momentum flux balance in the shock's its rest frame. This determines the incoming fluid velocity $v_0$ and the outgoing velocity $v_1$ in terms of the fractional increase $\Delta$ in the density across the shock. One finds  $v_0=\sqrt{(4+3 \Delta)/(4+\Delta)}/\sqrt{3}$ and $v_1=\sqrt{(4+\Delta)/( 4+3 \Delta)}/\sqrt{3}$.  Next, the rest-frame entropy density is directly related to the rest-frame energy density and is therefore enhanced behind the shock front by a factor of $(1+\Delta)^{3\over 4}$. Therefore, the outgoing entropy flux is enhanced relative to the incoming flux by $(1+\Delta)^{3\over 4} (\gamma_1 v_1)/(\gamma_0 v_0) = (1+\Delta)^{1\over 4} \sqrt{(4+\Delta)/(4+3\Delta)}\approx 1 +{1\over 64} \Delta^3$, for small $\Delta$. The entropy density behind the shock is larger than that in front by the same factor.

Entropy production results in the dissipation of shocks. Consider a sinusoidal density perturbation of initial amplitude $\epsilon$ which forms left- and right-moving shocks of strength $\Delta= \epsilon$. Averaging over space, the entropy deficit per unit volume is $-{3\over 64}\Delta^2 s_0$, where $s_0$ is the equilibrium entropy density. The rate of change of this deficit equals the rate at which the shocks generate entropy, which is ${1\over 64}c_s \Delta^3 s_0/\lambda_s$, where $\lambda_s$ is the mean shock separation. Hence, we obtain $\dot{\Delta}=-{1\over 6} (c_s/\lambda_s) \Delta^2$ so that shocks of amplitude $\epsilon$ decay in a time $t_d\sim 6 \lambda_s/(c_s \epsilon)$, larger than the shock formation time by a numerical factor (which, in our simplified model, is $\sqrt{3} \pi\approx 5$ in $d=3$). The shock amplitude decay introduces a short wavelength cutoff in the entropy deficit:
\be
s\approx s_0(1 -{3\over 32} \epsilon^2 \ln\left(\lambda_{2}/(C c_s \epsilon\, t)\right), 
\label{eq9}
\ee
with $C$ a constant (equal to ${1\over 6}$ in our simplified model). We have checked this picture in detailed numerical simulations in one, two and three dimensions. Fig. \ref{entropyfig} shows a full 3d numerical simulation compared with the prediction of Eq. (\ref{eq9}), with excellent agreement.

Not only do shocks generate entropy, shock-shock interactions generate vorticity, in a precisely calculable amount. For example, one can find a stationary solution representing two shocks intersecting on a line, leaving behind a ``slip sheet'' across which the tangential component of the velocity is discontinuous. In such steady-state flows the strength of the tangential discontinuity (and hence the vorticity) is proportional to $\Delta^3$ with $\Delta$ the shock amplitude. More generally, non-stationary configurations can generate parametrically larger vorticity and indeed, it is conceivable that in rare localized regions fully developed turbulence may occur. 

 Finally,  let us return to the production of gravitational waves from larger-amplitude perturbations such as have been invoked to explain the formation of black holes in the early universe. In second order perturbation theory, adiabatic perturbations with amplitude $\epsilon$ lead to a stochastic background of gravitational waves, produced at the Hubble radius, with spectral density $\Omega_g(f) h^2 \sim \epsilon^4 \Omega_\gamma h^2$ where $\Omega_\gamma h^2 \sim 4.2\times 10^{-5}$ is the fractional contribution of radiation to the critical density today~\cite{baumann}. As we shall show elsewhere~\cite{penturoklong}, shock collisions generate a parametrically similar contribution to the stochastic gravitational wave background, also on Hubble horizon scales. But because shocks form later, they emit gravitational waves at longer wavelengths, with frequencies which are lower by a factor of $\epsilon$.  In the scenario of Ref. \cite{Bird}, $30 M_{\odot}$ primordial black holes would form at a time $t\sim 10^{-4}$ seconds from high peaks in perturbations with rms amplitude $\epsilon\sim 10^{-1}$. At second order in perturbation theory these contribute a stochastic gravitational wave background with  $\Omega_g(f) h^2 \sim 4\times 10^{-9}$, at frequencies of $\sim 30$ nHz today. This is outside the exclusion window of the European Pulsar Timing Array, $\Omega_g(f) h^2 < 1.1 \times 10^{-9}$ at frequencies $f\sim 2.8$ nHz~\cite{EPTA}. However, for $\epsilon\sim 0.1$, the gravity wave background due to shocks peaks at $\sim 3$ nHz, inside the exclusion window, potentially ruling out the scenario of Ref.~\cite{Bird}. Gravitational wave detectors are now operating or planned over frequencies from nHz to tens of MHz (see, {\it e.g.}, Ref.~\cite{holom}), corresponding to gravitational waves emitted on the Hubble horizon at times from $10^{-4}$ to $10^{-30}$ seconds. In combination with detailed simulations of the nonlinear evolution of the cosmic fluid and consequent gravitational wave emission, these experiments will revolutionize our ability to constrain the physical conditions present in the primordial universe, an exciting prospect indeed.     
 
{\em Acknowledgments.} ---
We thank John Barrow, Dick Bond, Job Feldbrugge, Steffen Gielen, Luis Lehner, Jim Peebles, Dam Son and Ellen Zweibel for valuable discussions and correspondence. Research at Perimeter Institute is supported by the Government of Canada through Industry Canada and by the Province of Ontario through the Ministry of Research and Innovation. 



\begin{thebibliography}{99}
\bibitem{foottilt} We ignore the modest effect of scalar tilt on the power spectrum.
\end{thebibliography}
\end{document}